\documentclass{cernrep}
\usepackage{amsmath,amssymb,amsfonts,color,graphicx,cite}
\usepackage{hyperref}

\usepackage{ulem} 

\newcommand{\be}{\begin{equation}}
\newcommand{\ee}{\end{equation}}
\newcommand{\bea}{\begin{eqnarray}}
\newcommand{\eea}{\end{eqnarray}}
\newcommand{\One}{{\mathbf{1}}}

\newcommand{\eq}[1]{Eq.~\eqref{#1}}
\newcommand{\Figref}[1]{Fig.~\ref{#1}}

\newcommand{\tev}{\,\, \mathrm{TeV}}
\newcommand{\gev}{\,\, \mathrm{GeV}}

\title{Future Prospects for the Minimal Supersymmetric Standard Model}
\author{
Shehu~AbdusSalam~\footnote{\,\,\,\,email: abdussalam@sbu.ac.ir} \,\,$^1$, 
Safura~S.~Barzani$^1$,
Mohammadreza~Noormandipour$^{1,2}$
}
\institute{
  $^1$ Department of Physics, Shahid Beheshti University, Tehran, Islamic Republic of Iran \\
  $^2$ DAMTP, University of Cambridge, Wilberforce Road, Cambridge, CB3 0WA, UK
}

\begin{document}
\maketitle 
\begin{abstract}
Experimental collaborations for the large hadron collider conducted many and various searches for supersymmetry. In the absence of signals, lower limits were put on sparticle masses but usually within frameworks with (over-) simplifications relative to the entire indications by supersymmetry models. For complementing current interpretations of experimental bounds, we introduce a 30-parameters version of the R-parity conserving minimal supersymmetric standard model (MSSM-30). Using a sample of the MSSM-30 which are in harmony with cold dark matter, flavour and precision electroweak constraints, we explicitly show the prospects for assessing neutralino candidate dark matter in contrast to future searches for supersymmetry. The MSSM-30 parameters regions that are beyond reach to dark matter direct detection experiments could be probed by future hadron-hadron colliders.
\end{abstract}

\section{Introduction}
\label{sec:intro}
The Standard Model (SM) of particle physics successfully explains observed strong,
electromagnetic and weak interactions and is complete given the discovery of a neutral scalar particle near
$125 \gev$~\cite{Aad:2012tfa, Chatrchyan:2012xdj} at the large hadron collider (LHC). It however suffers the problem concerning weak scale naturalness -- the fact that the Higgs boson mass is not stable to radiative corrections. 
It also could not explain the dark matter relic abundance measured by the Planck experiment~\cite{Planck:2018vyg}.  
Supersymmetry (SUSY) provides a mathematically attractive theoretical frame for building models beyond the SM (BSM). The theory was built on a well-motivated extension of the Poincare symmetry in
quantum field theory by a new space-time symmetry with interesting phenomenological consequences. 
With SUSY, the presence of a fermionic partner is predicted for each SM bosonic
particle and vice versa.
 SUSY can be used to address the naturalness problem and for constructing models with credible dark matter (DM) candidates. 
The search for SUSY
signature is part of the aims of collider and dark matter search experiments.

The R-parity conserving minimal supersymmetric standard model (MSSM) has been analysed thoroughly via various phenomenological perspectives but there is not any evidence for SUSY from LHC and dark matter search experiments. Limits on the cross-sections for relevant SUSY
processes are published from the experiments. These were used for interpreting the experimental
results based on various SUSY models to derive implications. The interpretations made using simplified models~\cite{Alves:2011wf} and specific SUSY-breaking models push sparticle masses well into the TeV or multi-TeV regions.
The status for the SUSY-breaking parameters of complete models, compared to the simplified model scenarios, at the TeV scale are
  less finely interpreted mostly due to a large number of non-sensitive free parameters.

The experimental search for SUSY goes on. To continue searches beyond the LHC and current DM experiments, there are proposals for collider experiments with a higher centre of mass energies \cite{Schumann:2019eaa, Abada:2019ono}  and for upgrades of DM search experiments. The interesting question~\cite{AbdusSalam:2011hd} remains: Will the proposed experiments be able to discover or rule out the MSSM? 
Our take on this question is as follows. There is a
robust guiding principle for getting answers with the Bayesian global fits approach. With this approach, albeit computationally expensive, one may find prior-independent 
model features that could be decisively ruled out by experiments in the absence of discovery. Establishing the fate of the MSSM for experiments requires detailed explorations of the 100-dimensional parameters space. Most of the parameters slightly or not sensitive to collider measurements are strongly constrained by flavour physics or electric dipole moment (EDM) observables. Analyses taking into consideration all these classes of constraints via global fits usually lead to better and realistic conclusions. Experiences in support of this
   can be found within the phenomenological MSSM (pMSSM)~\cite{Djouadi:1998di, Profumo:2004at, AbdusSalam:2008uv, Berger:2008cq, AbdusSalam:2009qd,Conley:2010du, Arbey:2011un, Sekmen:2011cz, AbdusSalam:2012ir, CMS:2014mia, Besjes:2014lmc, Aad:2015baa, Marjanovic:2015tzv, Khachatryan:2016nvf, Fawcett:2016xoh}. Here we highlight one. 
   Just before LHC commissioning, there were expectations for a Higgs boson with mass ``around the corner'' of the large electron-positron limits. That was because of earlier conclusions from well-motivated analyses based on specific SUSY-breaking mechanisms (for instance, see~\cite{Buchmueller:2008qe, Buchmueller:2007zk, Ellis:2005tu}). The complementary global fit analyses of the pMSSM~\cite{AbdusSalam:2008uv, AbdusSalam:2009qd} showed that the contrary should be expected -- a multi-TeV scalar top-quark~\cite{AbdusSalam:2013qba} along with Higgs boson mass between 117 GeV and 128 GeV at 95\% Bayesian confidence region. We consider this to be a success for the pMSSM.  There is an extensive set of phenomenological studies and experiment limits based on the pMSSM in the literature. For instance, see~\cite{Djouadi:1998di, Profumo:2004at, AbdusSalam:2008uv, Berger:2008cq, AbdusSalam:2009qd,Conley:2010du, Arbey:2011un, Sekmen:2011cz, AbdusSalam:2012ir, CMS:2014mia, Besjes:2014lmc, Aad:2015baa, Marjanovic:2015tzv, Khachatryan:2016nvf, Fawcett:2016xoh} and there citations. 

Learning from the pMSSM, it is interesting to develop a broader MSSM phenomenological framework to complement the research directions based on (over-)simplified SUSY representations. An MSSM framework with 30 free parameters (MSSM-30) was
developed~\cite{AbdusSalam:2014uea} by systematically turning on CP- and flavour-violating parameters. From this cutting-edge SUSY framework, one can collect MSSM sample points that are compatible with flavour physics, EDM, and other state-of-the-arts low energy constraints. In this article, we propose that such MSSM samples could be the best and most credible to assess the MSSM with regards to current and future collider and DM experiments searching for SUSY.
 We analyse the
MSSM-30 posterior sample from the global fit reported in~\cite{AbdusSalam:2014uea} by simulating SUSY productions at 
100 TeV centre of mass proton-proton collision energy using the future circular collider software (FCCSW) \cite{fcc2}. This is a demonstrative example of what can be done for future collider studies since the sample of the MSSM points is from the year 2014 global fit with constraints that have changed substantially by now. In particular, the 10 fold improved limit on the electron EDM~\cite{ACME:2018yjb} is severely constraining. We expect the current EDM constraint to push the sparticle masses higher than what is presented here. Performing a new global fit is beyond the scope of this article. Using the ``old'' sample, we also analyse the current status and prospects for the MSSM-30 spectra to limits from the direct search for neutralino cold dark matter (CDM). In the next section, we review the construction of the 30-parameters MSSM. This is followed by the description 
of the simulations and analyses of the MSSM-30 sample and a conclusion in the last section.

\section{The 30-parameters MSSM}
\label{sec:mssm30}
The MSSM has about 100 free SUSY-breaking parameters. 
 The challenge of confronting these with data lead to
the development of various kinds of well-motivated SUSY-breaking models with few/several free parameters~\cite{Martin:1997ns, Chung:2003fi, Giudice:1998bp, Ibanez:2012zz}. 
 There is an approach for surveying parameters to cut down
their number by imposing minimal flavour violation (MFV)~\cite{Chivukula:1987fw, Buras:2000dm, DAmbrosio:2002vsn, Buras:2003jf, Hall:1990ac}
on the MSSM. The flavour-violating SUSY-breaking interactions are expressed in the basis which emphasises
their transformation properties under the approximate global flavour symmetries. 
As such we associate a power of the small symmetry-breaking size to the MSSM flavour-changing interactions in a way consistent with
the known flavour-changes of the SM. Counting the suppression by the global flavour symmetry-breaking parameter ranks their
size and hence gives a parameter-selection procedure, wherein one neglects all terms beyond a given fixed
order~\cite{Colangelo:2008qp}. This is the procedure used for developing~\cite{AbdusSalam:2014uea} the many-parameters MSSM: a 24-parameter
MSSM-24, which works at the lowest nontrivial order; a 30-parameter MSSM-30 which we address in this article; and
a 42-parameter, MSSM-42.

An important advantage of the procedure above is the ability to strengthen (or weaken) the MFV parameter-selection prescription at
will to exclude more (or fewer) flavour-violating interactions, simply by changing the order in the global flavour symmetry breaking
that is to be neglected. This is to be contrasted with the procedure of setting all off-diagonal mass terms to zero by hand
once and for all, as is done for the pMSSM. The systematic explorations of
these models should provide a better way for quantitative assessments of SUSY. 
Further, from a bottom-up perspective, choosing to work within a few-parameters framework has the potential that some physics that
may be important can be lost. For example, moving from the few-parameters constrained MSSM (CMSSM) to the 20-parameter pMSSM, as 
in~\cite{AbdusSalam:2008uv,AbdusSalam:2009qd}, changed the favoured masses of the Higgs boson and the scalar top-quark to 119-128 GeV
and order 3 TeV respectively. 

\subsection{The MSSM-30 construction}
\label{sec:mssm30const}
For completeness, we first provide a summary of the pMSSM and then the MSSM-30 construction. For both, the goal is to
develop the criterion for excluding flavour-changing and CP-violating interactions. This is achieved for the pMSSM by making
the following choices:
\begin{itemize}
\item $R$-parity conservation;
\item absence of flavour-violating interactions (when renormalised at TeV scales);
\item degenerate masses and negligible Yukawa couplings for the 1st/2nd generation sfermions;
\item absence of CP-violating interactions beyond those from the SM CKM matrix;
\item lightest neutralino should be the lightest SUSY particle (LSP) and should be a thermal relic.
\end{itemize}
This leads to a 19-parameters model which captures well the MSSM's lightest CP-even Higgs sector. However, some of these choices
are ultimately {\em ad-hoc} even if well-motivated. Some of the above selections were inspired by MFV considerations but none of
them is derived from it.

The MFV hypothesis~\cite{Chivukula:1987fw, Buras:2000dm, DAmbrosio:2002vsn, Buras:2003jf} relates the small size of flavour-violating effects to approximate symmetries. Accordingly, the starting point is to identify the large group, $G$, of flavour symmetries that the SM respects when all Yukawa couplings are set to
zero. It is based on the assumption that only spurion fields proportional to the SM Yukawa couplings break the symmetries, $G$.
Promoting the SM Yukawa couplings to spurions automatically generates the GIM~\cite{Glashow:1970gm} cancellations required by observations
once loop effects are included. Applying this to the MSSM will mean that low-scale MSSM flavour couplings can be reconstructed
entirely out of appropriate powers of the SM Yukawa coupling matrices, $Y_{U,D,E}$ and MSSM SUSY-breaking terms can be expanded as 
series of the $G$-invariant spurion factors~\cite{Ellis:2007kb,Colangelo:2008qp,Chivukula:1987fw, Buras:2000dm, DAmbrosio:2002vsn, Buras:2003jf,Hall:1990ac,Smith:2009hj}: 
\bea \label{massbreakingpars}
& &(M^2_Q)_{ij} = M^2_Q \left[ \delta_{ij} + b_1 (Y_U^\dagger
  Y_U)_{ij} + b_2 (Y_D^\dagger Y_D)_{ij} + c_1\{ (Y_D^\dagger Y_D
  Y_U^\dagger Y_U)_{ij} + H.c.\} + \ldots\right], \nonumber \\ 
& &(M^2_U)_{ij} = M^2_U \left[ \delta_{ij} + b_3 (Y_U
  Y_U^\dagger)_{ij} + \ldots \right],  \nonumber \\ 
& &(M^2_D)_{ij} = M^2_D \left[
  \delta_{ij} + [Y_D (b_6 + b_7 Y_U^\dagger Y_U) Y_D^\dagger]_{ij}
 + \ldots \right], \nonumber \\
& &(M^2_L)_{ij} = M^2_L \left[ \delta_{ij} + b_{13} (Y_E^\dagger
  Y_E)_{ij} + \ldots \right], \nonumber \\ 
& &(M^2_E)_{ij} = M^2_E \left[ \delta_{ij} +
  b_{14} (Y_E Y_E^\dagger)_{ij} + \ldots \right] \,,
\eea
and
\bea \label{ycouplingsbreakingpars}
& &(A^{'}_E)_{ij} = a_E \left[ \delta_{ij} + b_{15} (Y_E^\dagger
  Y_E)_{ij} + \ldots \right], \nonumber\\ 
& &(A^{'}_U)_{ij} = a_U \left[ \delta_{ij} + b_9 (Y_U^\dagger
  Y_U)_{ij} + b_{10} (Y_D^\dagger Y_D)_{ij} + \ldots \right], \nonumber \\
& &(A^{'}_D)_{ij} = a_D \left[ \delta_{ij} + b_{11} (Y_U^\dagger
  Y_U)_{ij} + b_{12} (Y_D^\dagger Y_D)_{ij} + c_6 (Y_D^\dagger Y_D
  Y_U^\dagger Y_U)_{ij} + \ldots \right].
\eea
Here the MSSM trilinear scalar couplings take the form $(A_{E,U,D})_{ij} = (A^{'}_{E,U,D} Y_{E,U,D})_{ij}$ and $i, j$ run over the family
indices $1,2,3$. These soft SUSY-breaking parameters are set at the TeV scale. Since this is the energy scale which the experiments under
consideration seek to probe, we avoid setting the parameters at higher energy scales such as the gauge couplings unification scale. 
The series is not infinite due to the Cayley-Hamilton identities for $3 \times 3$ matrices. Therefore there will be a finite number
of the coefficients $b_k$ and $c_k$ (finite $k = 1, 2, 3, 4, \ldots$).
Generically, for such series expansions of the matrices, the values which with the coefficients can
take will span many orders of magnitude. This is where the power of the MFV hypothesis lies~\cite{DAmbrosio:2002vsn, Buras:2003jf, Hall:1990ac, Colangelo:2008qp}. It stipulates that all the $b_k$ and $c_k$ 
must be of order unity. That way, all small numbers suppressing flavour changes will solely come from Yukawa matrices factors.
Now, ad-hocly setting $b_k = c_k = 0$ will be a crude way to truncate the MSSM parameters to a flavour-blind set with 14 free 
parameters. The sfermion masses within each family will be degenerate. Alleviating the degeneracy to only the 1st/2nd generations
then gives the 19-parameter pMSSM. Instead of the truncation by hand, the MSSM parameters can be reduced by using the approximate
global flavour symmetry by dropping terms smaller than a particular fixed order in small mixing angles (like the Cabibbo angle). 

Such a systematic approach for selecting MSSM parameters have been prescribed in~\cite{Colangelo:2008qp}. The counting rule makes use of 
the hierarchical structure along the off-diagonal terms of the Yukawa matrices in terms of the Cabibbo angle,
$\lambda = \sin\theta_{CB} \simeq 0.23$ to make a new complete basis vectors with a closed algebra under multiplication:
\begin{eqnarray}
  \label{xbase}
  \begin{array}{cccc}
    X_1 = \delta_{3i} \delta_{3j}, & X_2 = \delta_{2i} \delta_{2j}, & X_3 = \delta_{3i} \delta_{2j}, & X_4 = \delta_{2i} \delta_{3j},\\
    X_5 = \delta_{3i} V_{3j}, & X_6 = \delta_{2i} V_{2j}, & X_7 = \delta_{3i} V_{2j}, &X_8 = \delta_{2i} V_{3j}, \\
    X_9 = V^*_{3i} \delta_{3j}, & X_{10} = V^*_{2i} \delta_{2j}, & X_{11} = V^*_{3i} \delta_{2j}, & X_{12} = V^*_{2i} \delta_{3j},\\
    X_{13} = V^*_{3i} V_{3j}, & X_{14} = V^*_{2i} V_{2j}, & X_{15} = V^*_{3i} V_{2j}, & X_{16} = V^*_{2i} V_{3j}.
  \end{array}
\end{eqnarray}
The basis vectors are all of order one since each has at least one entry of order unity. This way each of the MFV parameters, in the $X_k$
basis, can be associated with an order in $\lambda$ depending on what coefficients come with it in the soft SUSY-breaking terms. And
once the accuracy of calculations is chosen in the form ${\cal O}(\lambda^n),\, n = 1,2,\ldots$, then some of the terms in the
SUSY-breaking mass or couplings can be systematically discarded from the expansion expressed in the $X_k$ basis.

The prescription above stems from the observation that in \eq{massbreakingpars} and \eq{ycouplingsbreakingpars} after the
reduction of the infinite series into few terms via Cayley-Hamilton identities, large pieces of terms such as $(Y_U^\dagger\, Y_U)_{ij}^2$
and $(Y_U^\dagger \,Y_U)_{ij}$ are proportional to $V_{3i}^* \,V_{3j}$ where $V$ is the CKM matrix. Next to this large pieces, relatively
smaller terms proportional to $V_{2i}^* \,V_{2j}$ follows. So in the SUSY-breaking mass or trilinear coupling expressions, 
$V_{3i}^* \,V_{3j}$ and $V_{2i}^* \,V_{2j}$ can be used as basis vectors which appear with coefficients of order one
and $y_c^2 \sim \lambda^8$ respectively, instead of $(Y_U^\dagger \,Y_U)_{ij}^2$ and $(Y_U^\dagger \, Y_U)_{ij}$. Similarly, 
$(Y_D^\dagger \,Y_D)_{ij}$ and $(Y_D^\dagger \, Y_D)_{ij}^2$ can be replaced by $\delta_{i3}^* \, \delta_{j3}$ and
$\delta_{i2}^* \, \delta_{j2}$ with order $y_b^2$ and $y_s^2$ coefficients respectively. Here $\delta_{ij}$ represents the unit matrix
in family space. This procedure leads to a 42-parameter model, MSSM-42~\cite{AbdusSalam:2014uea},
by dropping order $\lambda^6 \sim 10^{-4}$ terms
from the soft SUSY-breaking masses and couplings in \eq{massbreakingpars} and \eq{ycouplingsbreakingpars}. The complete
MSSM parameters for this are: 
\bea \label{mfvparsx}
\widetilde{M}_1 &=& e^{\phi_1} M_1, \,\quad  \widetilde{M}_2=e^{\phi_2} M_2, \,\quad M_3, \,\quad \widetilde{\mu}=e^{\phi_\mu} \mu, \,\quad M_A, \,\quad \tan \beta, \nonumber \\ 
M^2_Q &=& \tilde{a}_1\ \One + x_1 X_{13} + y_1 X_1 + y_2 X_5 + y^*_2 X_9,  \nonumber \\
 M^2_U &=& \tilde{a}_2\ \One  + x_2 X_1, \nonumber \\
 M^2_D &=& \tilde{a}_3\ \One + y_3 X_1 + w_1 X_3 + w^*_1 X_4, \nonumber \\
 M^2_L &=& \tilde{a}_6\ \One + y_6 X_1, \nonumber \\
 M^2_E &=& \tilde{a}_7\ \One + y_7 X_1, \nonumber \\
 A_E &=& \tilde{a}_8 X_1 + w_5 X_2, \nonumber \\
 A_U &=& \tilde{a}_4 X_5 + y_4 X_1 + w_2 X_6, \nonumber \\
 A_D &=& \tilde{a}_5 X_1 + y_5 X_5 + w_3 X_2 + w_4 X_4.
\eea
Since the SUSY-breaking mass parameters are Hermitian then $\tilde{a}_{1-3,6,7} > 0$, and $x_1, x_2, y_1, y_3, y_6, y_7$ must be real
while the other coefficients can be complex. Here $M_1, M_2, M_3$ are gaugino mass parameters with the first two allowed to be complex
by turning on the CP-violating phases $\phi_1, \phi_2$. $\phi_\mu, \mu, M_A, \tan \beta$ are the MSSM Higgs/Higgsino sector parameters. 

Working to order ${\cal O}(\lambda^4) \sim {\cal O}(10^{-3})$, only $x_{1-2}, y_1, y_3, y_6, y_7 \in \mathbb{R}$ and
$\tilde{a}_{4,5,8}, y_{4-5} \in \mathbb{C}$ from \eq{mfvparsx} remain, making the MSSM-30 with the following parametrisation:
\bea \label{mfvpar30}
 & &\widetilde{M}_1 = e^{\phi_1} M_1, \,\quad  \widetilde{M}_2=e^{\phi_2} M_2, \,\quad M_3, \,\quad \widetilde{\mu}=e^{\phi_\mu} \mu, \,\quad M_A, \,\quad \tan \beta, \nonumber \\ 
& &M^2_Q = \tilde{a}_1\ \One + x_1 X_{13} + y_1 X_1, \nonumber\\ 
& &M^2_U = \tilde{a}_2\ \One  + x_2 X_1, \nonumber\\
& &M^2_D = \tilde{a}_3\ \One + y_3 X_1, \nonumber\\
& &M^2_L = \tilde{a}_6\ \One + y_6 X_1, \nonumber\\
& &M^2_E = \tilde{a}_7\ \One + y_7 X_1, \nonumber\\
& &A_E = \tilde{a}_8 X_1, \nonumber\\
& &A_U = \tilde{a}_4 X_5 + y_4 X_1, \nonumber\\
& &A_D = \tilde{a}_5 X_1 + y_5 X_5. 
\eea

\begin{figure}[!ht]
  \begin{tabular}{c}
    \includegraphics[angle=0, width=.9\textwidth]{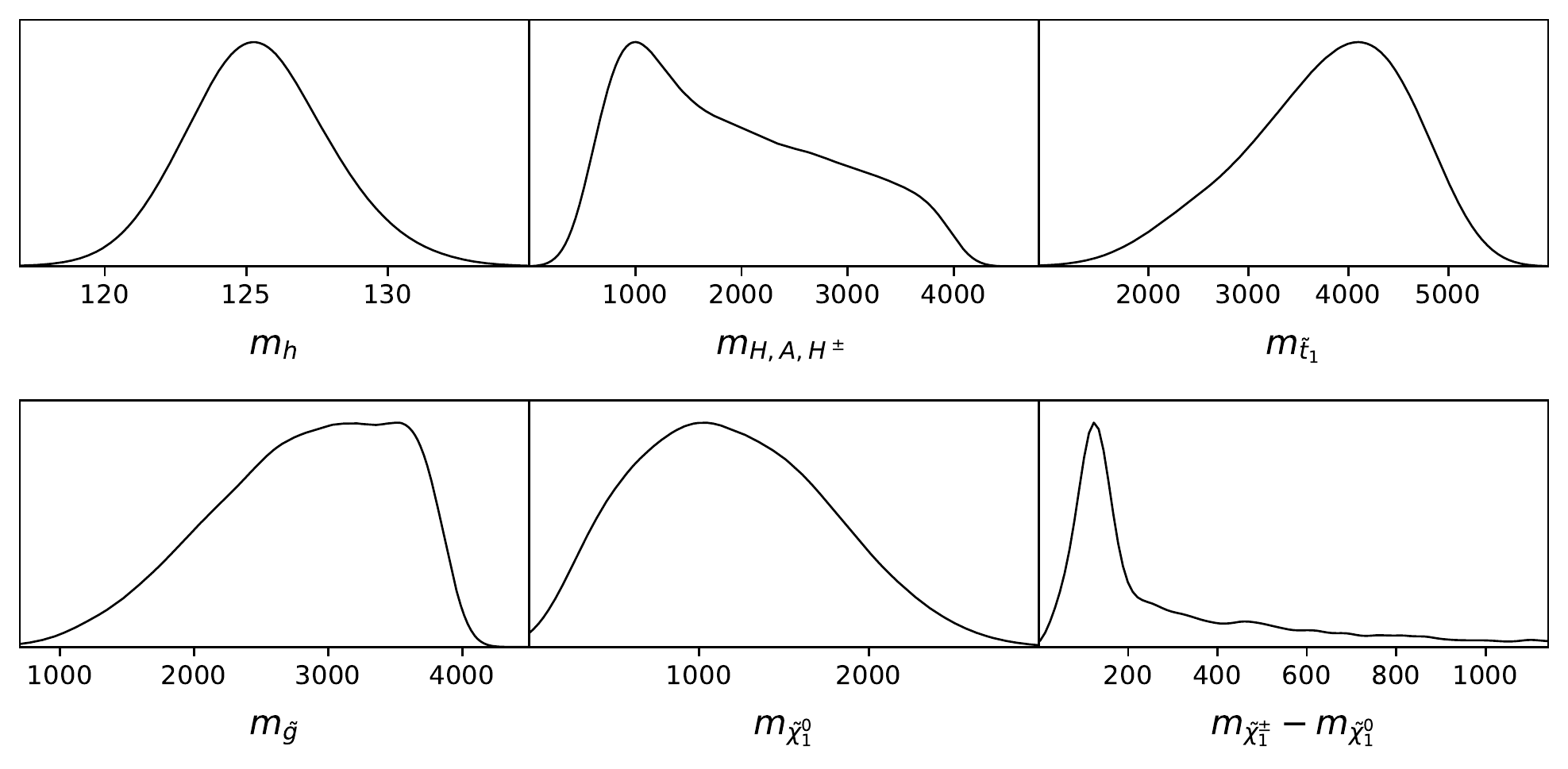}
  \end{tabular}
  \caption{The MSSM-30 1D posterior distribution for Higgs bosons' ($m_{h,H,AH^{\pm}}$), scalar-top quark ($m_{\tilde{t}}$), 
    gluino ($m_{\tilde{g}}$), and neutralino ($m_{\tilde{\chi}_1^0}$) masses. The last plot is the distribution of the
     lightest chargino-neutralino mass difference ($m_{\tilde{\chi}_1^{\pm}} - m_{\tilde{\chi}_1^0}$). All masses here are in $\gev$ unit.}
  \label{massdistribution}
\end{figure}

A global fit of the MSSM-30 to low energy data was reported in~\cite{AbdusSalam:2014uea}. As can be seen in~\Figref{massdistribution},
the bulk of the posterior distributions have sparticle masses too heavy, and possibly beyond reach at the LHC~\footnote{Posterior plots in this article were made using \texttt{GetDist} \cite{Lewis:2019xzd}. The lightest Higgs boson mass distribution shows the Gaussian distribution as used for constraining the MSSM-30 parameters.}

\section{The MSSM-30 sample as benchmarks for future collider and CDM search studies} \label{anal}
By using the MSSM-30 sample, various posterior distributions for parameters/variables relevant for future collider design or discovery-reach analyses can be computed. The posterior probability density, $P(\underline \theta |\underline d)$, which is the result of the Bayesian global fit of the MSSM-30 to data, can be used to obtain the posterior of any function, $f(\underline{\theta})$:
\bea \label{derived_post}
P(f|\underline{d}) = \int{P(f,\underline{\theta} | \underline{d}) d \underline{\theta}} = \int{P(f|\underline{\theta}, \underline{d}) P(\underline{\theta} | \underline{d} ) d \underline{\theta} } = \int{\delta(f(\underline{\theta}) - f) P(\underline{\theta} | \underline{d} ) d\underline{\theta} } 
\eea
where $\delta$ is the Dirac delta function. Examples of $f(\underline{\theta})$ is the cross section for SUSY production at 100 TeV via Monte Carlo simulations and the neutralino-nucleons scattering cross section relevant for dark matter direct detection searches. The posterior distributions for low-energy physics observables can be directly compared with the cross sections. 
Ultimately, one may want to decide whether not to perform MSSM-related collider studies by using spectra known to be ruled out, say by some low-energy physics observable.  In this section,  we compute the cross-sections for the MSSM-30 (i) sparticles production at $100 \tev$ centre of mass energy of a future circular proton-proton collider, and (ii) neutralino-nucleon scattering for comparison with the limits from DM direct detection experiments. These are then used to show how collider searches for SUSY could complement the DM direct detection search experiments.

\subsection{MSSM-30 SUSY productions at $100 \tev$}
Retrospectively, consider the physics performance analyses in~\cite{ATLAS:1999vwa} and~\cite{Ball:2007zza}. These SUSY studies were done using the minimal supergravity framework. Perhaps, those should have come along with other complementary frameworks such as the pMSSM and the MSSM-30. Now, extrapolating into the future, for analysing sensitivities of proposed experiments, it is crucial to systematically develop new MSSM frameworks at the relevant energy scales.
To optimise the future circular collider performance and precision studies,
MSSM-30 benchmarks should be used to complement current efforts which are mostly based on a simplified models approach. 
In this section, we compute the MSSM-30 SUSY-production cross-section at 100 TeV proton-proton collisions. In section~\ref{mssmcdmlimits}, we put the MSSM-30 sample within the context of DM direction detection limits.
Doing this reveals the MSSM-30 regions that cannot be probed by the considered class of DM direct detection experiments but which could be covered at the proposed hadron collider.

\begin{figure}[!t]
  \begin{tabular}{ll}
    (a) \includegraphics[angle=0, width=.45\textwidth]{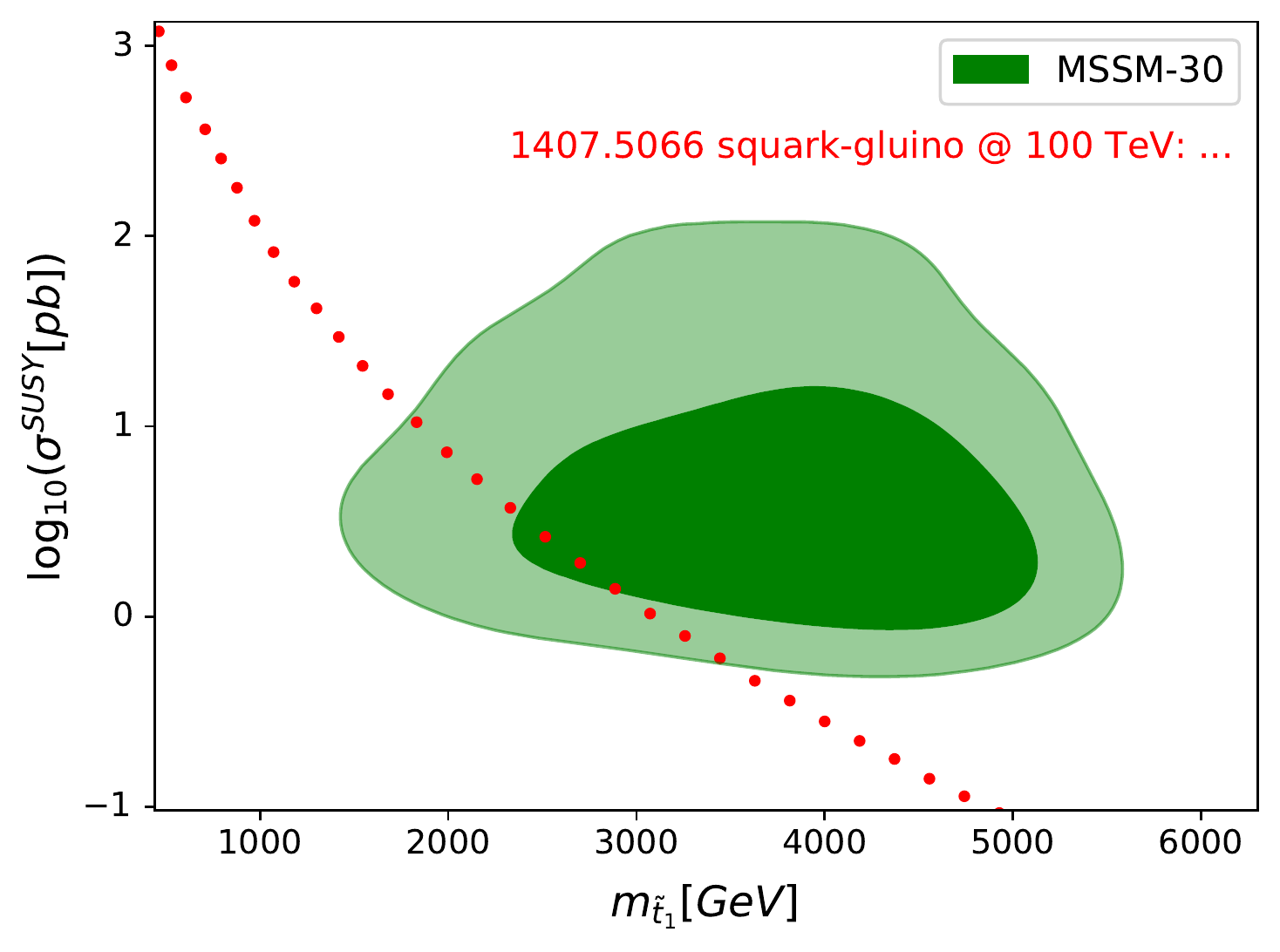}
    (b) \includegraphics[angle=0, width=.45\textwidth]{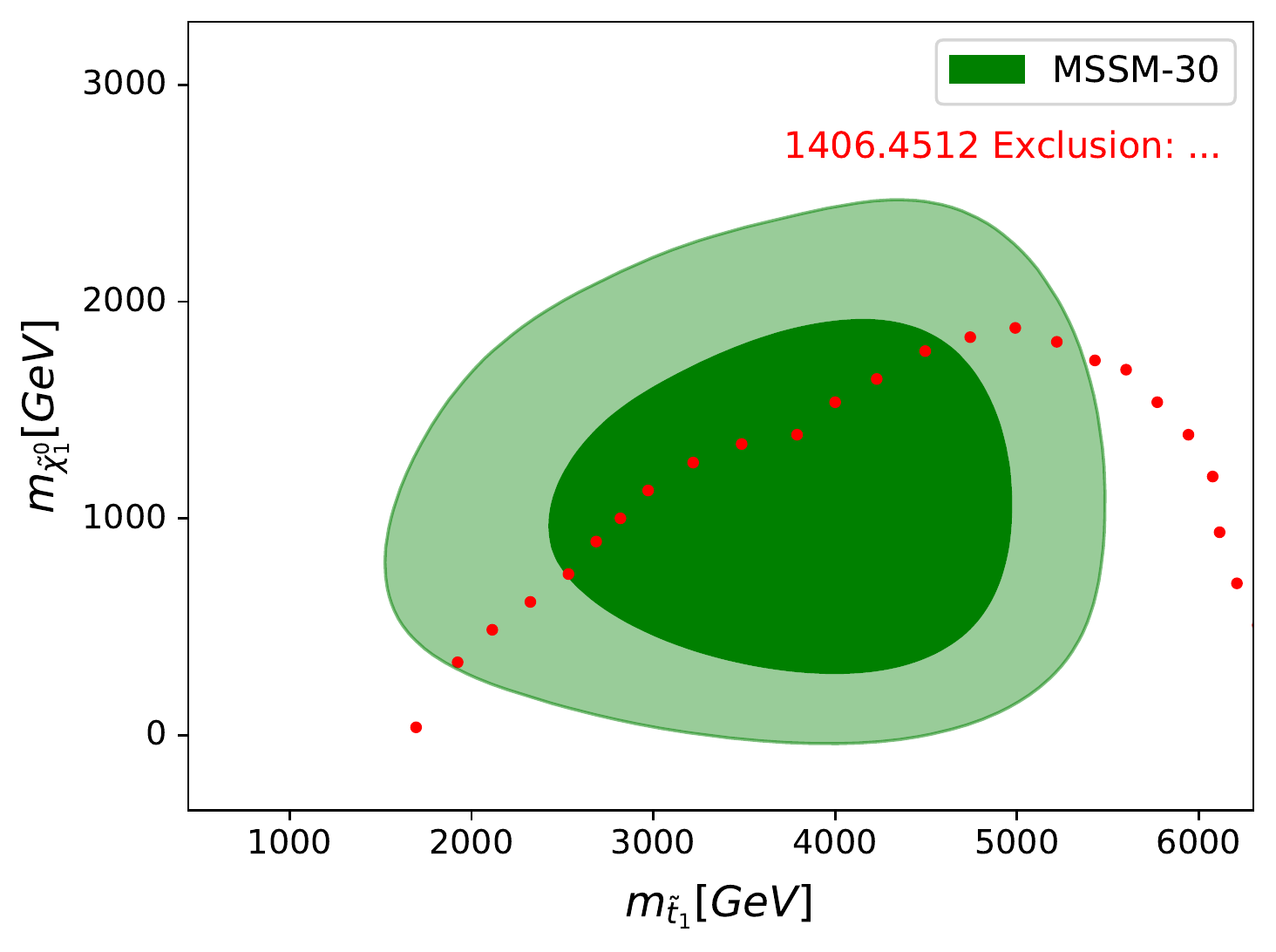} 
  \end{tabular}
  \caption{2D posterior distributions for the MSSM-30, {\bf (a)} all SUSY production at $\sqrt s = 100 \tev$ proton-proton collider versus the lightest top-squark mass compared to the simplified model mass-degenerate squark-gluino production \cite{Borschensky:2014cia}; and {\bf (b)} top-squark versus neutralino mass in comparison to the simplified model discovery reach contour line from \cite{Cohen:2014hxa}.}
  \label{fg.sigmas}
\end{figure}

For computing the MSSM-30 total SUSY production cross sections at $\sqrt s = 100 \tev$ proton-proton collider, the sample of parameter points in the SUSY Les Houches Accord (SLHA) format were passed through the FCCSW~\cite{fcc2}.
It uses \texttt{Pythia8} \cite{Sjostrand:2014zea} for generating 10000 Monte Carlo events for each MSSM-30 point. 
  In \Figref{fg.sigmas} we show the two-dimensional posterior distribution for (a) the SUSY production cross sections at leading order (LO) against the lightest top-squark mass, and (b) the top-squark versus neutralino masses. For both plots, the inner and outer contours enclose 68$\%$ and 95$\%$ Bayesian credibility regions, respectively, based on the low-energy physics measurements. 

  In \Figref{fg.sigmas}(a), the MSSM-30 LO cross section distribution is compared with the squark-gluino production cross sections at 100 TeV proton-proton collider computed based on simplified SUSY models (SSM) \cite{Borschensky:2014cia}. The latter is represented by the dotted contour line~\footnote{The contour points were extracted using \texttt{EasyNData}~\cite{Uwer:2007rs} from Fig.4 of \cite{Borschensky:2014cia}. Note that computing the MSSM-30 cross sections at the next-to leading order and next-to leading logarithmic precisions, as done in \cite{Borschensky:2014cia}, will not change our conclusions.}. 
This comparison explicitly shows the MSSM-30 region that will be missed should SUSY be represented by the SSM scenario considered. 
The 95$\%$ Bayesian credibility region of the MSSM-30 sample, as shown in \Figref{fg.sigmas}(a) is very different in contrast with the simplified model expectation. 
\Figref{fg.sigmas}(b) shows the two-dimensional posterior distribution for the MSSM-30 top-squark versus neutralino masses compared to the top-squark exclusion sensitivities at 100 TeV using SSM (Fig.5 of reference~\cite{Cohen:2014hxa}). According to this, there are indeed MSSM-30 points that cannot be excluded. 

\subsection{MSSM-30 neutralino-nucleon cross sections} \label{mssmcdmlimits}
\begin{figure}[!t] 
  \centering 
  (a) \includegraphics[angle=0, width=.45\textwidth]{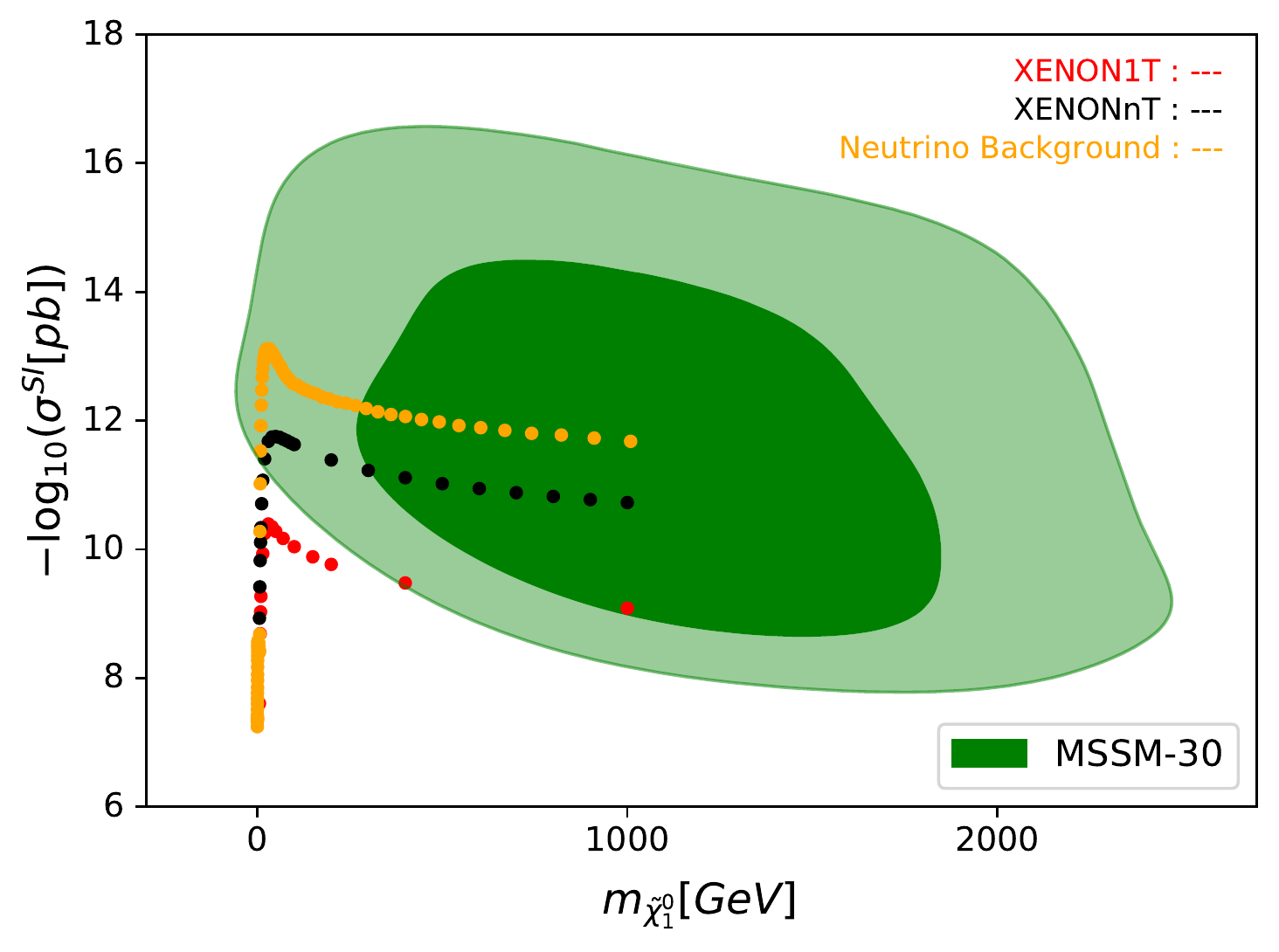}
  (b) \includegraphics[angle=0, width=.45\textwidth]{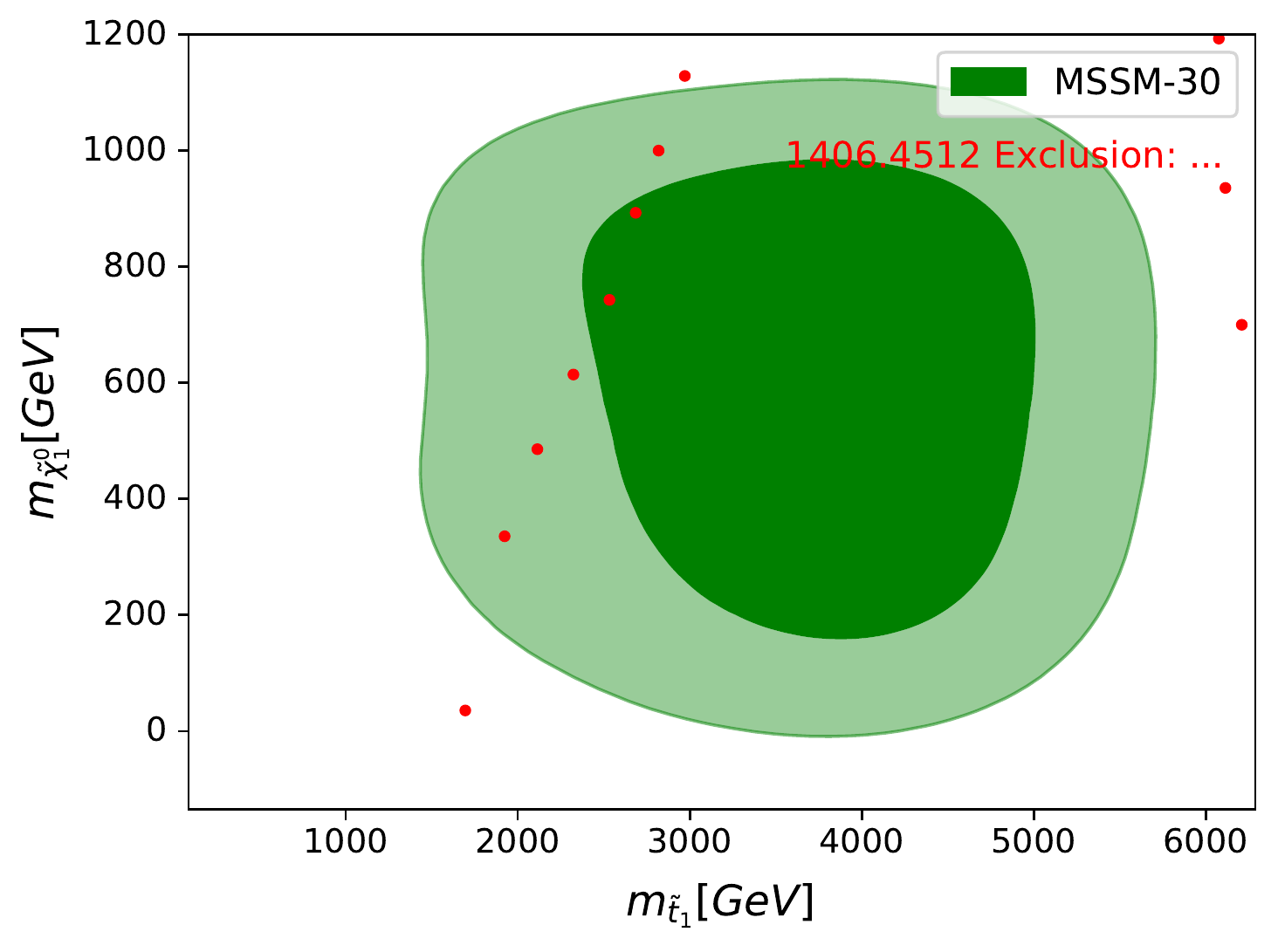}     
  \caption{{\bf (a)} The MSSM-30 posterior distribution of the neutralino mass versus the neutralino-nucleons spin-independent
    scattering cross-section. The dotted curves represent upper limits from DM direct search experiment XENON1T and its proposed future projection XENONnT. The ``Neutrino Background'' curve shows the cross section values below which DM candidate particles and background neutrinos may not be separable. As such the MSSM-30 sample points above this curve cannot be probed by DM direct detection experiments. {\bf (b)} The top-squark versus neutralino masses in comparison with the simplified model discovery
      reach contour line from \cite{Cohen:2014hxa} for the MSSM-30 points that are buried under the dark matter direct
      searches neutrino-background floor or which have neutralino LSP mass greater than 1 TeV. For both plots, the inner and outer contours enclose 68$\%$ and 95$\%$ Bayesian credibility regions respectively.  }  
  \label{fig.sigmas}
\end{figure}

The fact that no DM particle is detected is 
 used to put limits on the neutralino-nucleons spin-independent
cross section based on some assumed local neighbourhood DM 
density and velocity distribution (see e.g. \cite{Schumann:2019eaa}). The spin-independent
neutralino-nucleus elastic scattering cross section is given by  
\be
\sigma \approx \frac{4 m^2_{\tilde \chi^0_1} m^2_{T}}{\pi (m_{\tilde
    \chi^0_1} + m_T)^2} [Z f_p + (A-Z) f_n]^2,
\ee
where $m_T$ is the mass of the target nucleus and $Z$ and $A$ are the
atomic number and atomic mass of the nucleus, respectively.  $f_p$ and
$f_n$ are neutralino couplings to protons and neutrons, given
by~\cite{Gelmini:1990je,Drees:1992rr,Drees:1993bu,Jungman:1995df,Ellis:2000ds}
\be 
f_{p,n} = \sum_{q=u,d,s} f^{(p,n)}_{T_q} a_q \frac{m_{p,n}}{m_q} +
\frac{2}{27} f^{(p,n)}_{TG} \sum_{q=c,b,t} a_q  \frac{m_{p,n}}{m_q}, 
\label{quds}
\ee
where $a_q$ are the neutralino-quark couplings and $f^{(p,n)}_{T_q}$ denote the quark
content of the nucleon. For the MSSM-30 posterior sample points that pass the limits from LUX \cite{Akerib:2015rjg}, PANDA \cite{Cui:2017nnn} and XENON \cite{Aprile:2018dbl} experiments, the 2D posterior distribution 
 of the neutralino mass, $m_{\tilde{\chi}^0_1}$, versus its scattering cross section
with nucleons (here taken as the average of the neutralino scattering with proton and neutron computed
  using \texttt{micrOMEGAs} \cite{Belanger:2008sj}), $\sigma^{SI}$,
is shown in \Figref{fig.sigmas}(a). The contour lines show how
XENON1T limit could marginally probe the MSSM-30 sample. The XENONnT future projection has an improved sensitivity to the sample 
 but still cannot assess the entire sample.
This is because the MSSM-30 lightest neutralinos are strongly Higgsino-like with suppressed cross-sections to the nucleons. 
There is a significant region, beyond the ``neutrino floor'', which will escape the direct detection experiments.
Interestingly, most of the points within this region can be reached by 100 TeV hadron-hadron collider based on the contour line,
  from \cite{Cohen:2014hxa}, shown in \Figref{fig.sigmas}(b). 

\section{Conclusion}
\label{sec:concl} 
In~\cite{AbdusSalam:2014uea} a 30-parameters MSSM frame, MSSM-30, was developed by using symmetry arguments and MFV hypothesis. The MSSM-30 was then fit globally to low-energy observable related to the Higgs boson mass, the electroweak physics, B-physics, the electric dipole moments of the leptons and DM relic density.  
In this article, we highlight the importance of the fact that deriving robust conclusions about the MSSM with respect to proposed future experiments necessarily require interpretations and phenomenological studies within systematically constructed frameworks. We have used a posterior sample from a global fit of the MSSM-30 which features multi-TeV squarks and gluino masses. The neutralinos are also heavy with masses, $m_{\chi} \sim \textrm{1 to 2 TeV}$. The sample is analysed against the limits from experiments looking for DM particles. We show that the proposed XENONnT's direct search for neutralino DM will probe a significant portion of the MSSM parameter space. However, there are MSSM regions buried under the irreducible neutrino backgrounds for such a class of experiments. The searches for SUSY at proposed future colliders with higher energies can complement the neutralino DM direct detection in probing the neutrino backgrounds-buried MSSM regions.

Since the MSSM-30 global fit in~\cite{AbdusSalam:2014uea}, most of the electroweak precision, flavour physics and EDM observables have changed significantly. So the conclusion here should be considered solely based on the MSSM-30 sample used, and for demonstrating what could be done for using the framework for future colliders. There is a need to make a new global fit of the MSSM-30 with up to date constraints. We expect that this will not change the conclusion of this article.

\paragraph*{Acknowledgements:} For discussions and assistance in enabling of codes, we would like to thank 
Michele Silviaggi, Clement Helsens, Collin Bernet, Javad Ebadi, Leila Kalhor, Peter Skands and Werner Porod.

\end{document}